\documentclass[aps,prb,twocolumn,groupedaddress,showpacs,amssymb,amsmath]{revtex4}
\usepackage{graphicx}

\begin{document}

\title{Early stages of radiation damage in graphite and carbon nanostructures: \\
 A first-principles molecular dynamics study}

\author{Oleg V. Yazyev}
\email[Electronic address: ]{oleg.yazyev@epfl.ch}
\author{Ivano Tavernelli}
\author{Ursula Rothlisberger}
\author{Lothar Helm}
\affiliation{Ecole Polytechnique F\'ed\'erale de Lausanne (EPFL), \\
            Institute of Chemical Sciences and Engineering, \\
            CH-1015 Lausanne, Switzerland}

\date{\today}

\begin{abstract}

Understanding radiation-induced defect formation in carbon
materials is crucial for nuclear technology and for the manufacturing of
nanostructures with desired properties. Using first principles molecular
dynamics, we perform a systematic study of the non-equilibrium processes 
of radiation damage in graphite. 
Our study reveals a rich variety of defect structures (vacancies, interstitials,
intimate interstitial-vacancy pairs, and in-plane topological defects) 
with formation energies of 5--15~eV. 
We clarify the mechanisms underlying their creation and find
unexpected preferences for particular structures. 
Possibilities of controlled defect-assisted engineering of nanostructures 
are analyzed. In particular, we conclude that 
the selective creation of two distinct low-energy intimate Frenkel pair defects
can be achieved by using a 90--110~keV electron beam irradiation.

\end{abstract}


\maketitle

\section{INTRODUCTION}

Radiation resistance of graphite has been one of the major concerns
of the nuclear industry.\cite{Simmons65,Telling03} 
Nowadays, radiation treatment by high-energy 
electrons or ions is also viewed as a versatile tool for the design of 
new materials. The formation of irradiation-induced defects in graphite-like 
layered carbon nanostructures (multiwalled and bundled carbon 
nanotubes, nanoonions, etc.) 
changes their mechanical \cite{Kis04} and electronic 
properties \cite{Miko03,Han03} and may even trigger 
dramatic structural changes.\cite{Banhart97,Terrones00} 
However, the structure and dynamics of defects in 
graphite and carbon nanostructures as well as the mechanisms underlying 
their creation and transformation remain elusive. 
This knowledge is crucial for a defect-assisted engineering of 
nanostructures with applications in, e.g., 
manufacturing of nanoelectromechanical systems.\cite{NEMS}       

Radiation damage of matter is governed by the displacement of atoms 
from their equilibrium positions due to electronic excitations and direct
collisions of high-energy particles with the nuclei. In metals and narrow 
band gap semiconductors electronic excitations quench instantaneously, 
leaving collisions with nuclei as the sole mechanism responsible for 
the creation of defects in graphite and related carbon materials.\cite{Banhart99} 
If the kinetic energy transferred from a high-energy electron or ion to the nucleus
is higher than the displacement threshold $T_d$, a carbon atom can leave 
its initial position to form a metastable defect structure on 
a sub-picosecond timescale. Such events are called knock-on displacements.
For highly anisotropic layered carbon materials the threshold of the off-plane
displacement is $T^\bot_d$$\approx$15-20~eV \cite{Banhart97} 
while a creation of defect due to the in-plane knock-on collision requires higher transferred energies, $T^{||}_d$$\geq$30~eV. 
Possible defects produced by radiation damage include separated and intimate \cite{Ewels03} pairs of interstitial atoms and vacancies,
and in-plane topological defects involving non-sixmembered rings, e.g.
Stone-Wales defect.\cite{Stone86,Kaxiras88}
The existence of defects in carbon nanostructures has been
confirmed by direct observations.\cite{Hashimoto04,Urita05}

Upon knock-on events a large amount of energy is transferred to 
only a few degrees of freedom. The resulting defect structures 
formed on a picosecond timescale depend on the magnitude and on 
the direction of the transferred momentum and determine the fate
of the system at longer timescales. Therefore gaining control 
over the early stages of defect formation by tuning the irradiation
conditions will make the paradigm of the defect-assisted engineering
feasible. Molecular dynamics (MD) simulations performed with empirical 
potentials \cite{Nordlund96} or tight-binding models \cite{Crespi96,Ajayan98,Krasheninnikov05} 
have been used for the studies of radiation damage of various carbon materials. 

In this work, we report a {\it systematic first principles} study of 
the early stages of radiation damage of graphite, 
a general model for closely related layered carbon nanostructures.
The paper is organized in the following way: In Section~\ref{sec:methods}
we provide a description of the computational methods used in this work.
The observed defect structures, mechanisms of their formation, and 
practical implications are discussed in Section~\ref{sec:results}.
Section~\ref{sec:conclusions} briefly concludes our work.

\section{\label{sec:methods}COMPUTATIONAL METHODS}

\begin{figure}
\includegraphics[width=8.6cm]{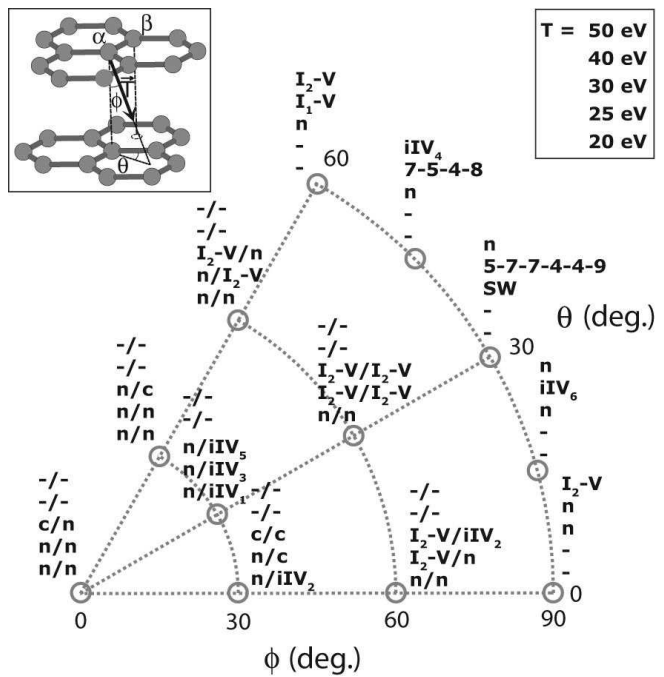}
\caption{\label{fig:char1}  
Polar coordinates
representation of the simulation outcomes as a function of the parameters $T$, $\phi$, and $\theta$.
The defect structures at each parameter set are given according to the nomenclature shown in Figs.~\ref{fig:eps1} 
and \ref{fig:eps2}. 
Other labels correspond to:
'n' -- no defect formation,
'c' -- displacement cascade, and '-' -- simulation not performed.  
The different outcomes corresponding to the off-plane displacements of $\alpha$/$\beta$ carbon atoms 
(see left inset) are shown separately. 
For each pair of parameters ($\phi$, $\theta$) the outcomes at different values of $T$
are listed according to the values given in the right inset.  
The left inset shows the definition of the parameters determining the knock-on displacements.
}
\end{figure}

By using \textit{ab initio} molecular dynamics 
we simulate the process of defect formation after the initial transfer
of a momentum $\vec{T}$ to one of the carbon atoms in the system.
The periodic model system consists of a unit cell with 108
carbon atoms, which contains two graphene sheets with stacking ABAB.
The dimension of the unit cell in the direction perpendicular to
the graphene planes was fixed to 6.7~\AA\ in accordance with   
the experimental inter-layer distance 3.35~\AA.\cite{Hanfland89}
This distance shows only weak variation among different layered 
carbon nanostructures. Our computational methodology 
is based on density functional theory (DFT), which lacks a correct
description of weak van der Waals interactions between graphene planes.
However, by fixing the unit cell dimension in the direction perpendicular to
the graphene planes we provide a realistic description of layered carbon 
nanostructures without any explicit inclusion of van der Waals 
forces. 
The in-plane distance between two periodic images is 12.7~\AA\,
which is large enough to ensure localization of the defect within
the unit cell. 
A coarse sampling of the irreducible wedge of the 
space spanned by the magnitude of transferred energy 
$T$ and the pair of angles $\phi$$\in$$[0^\circ; 90^\circ]$ 
and $\theta$$\in$$[0^\circ; 60^\circ]$ (Fig.~\ref{fig:char1}, inset) 
has been performed. 

The \textit{ab initio} MD simulations were
carried out using the \texttt{CPMD} plane wave DFT code\cite{CPMD} 
and the Perdew, Burke, and Ernzerhof exchange-correlation density
functional.\cite{Perdew96} A plane wave kinetic energy cutoff of 60~Ry and 
norm-conserving pseudopotentials \cite{Troullier91} have been used. 
The simulations were performed within the spin-unrestricted 
formulation of DFT starting from an initial guess
asymmetric with respect to the spin components. Such a starting configuration
is required in order to ensure
a broken-symmetry path of bond breaking events.\cite{Gunnarsson76} 
The first 100~fs of each MD simulation
were performed using the Born-Oppenheimer scheme.
The MD timestep was set to 0.5~fs.
In our simulations we observed that during the first 100~fs the transferred
kinetic energy was well dissipated over the entire system.  
The initial simulation was followed by a Car-Parrinello simulation \cite{Car85}
carried out using a Nos\'e-Hoover thermostat \cite{Nose84a} 
(350~K) until a stable defect structure was reached (about 1~ps).
This thermal coupling methodology provides a realistic 
description of the excess kinetic energy dissipation after knock-on
collisions of reasonably low transferred energies.
The Car-Parrinello equations of motion were integrated with a
time step of 0.1~fs using a fictitious electron mass
of 400~a.u.
Finally, the obtained defect structures were relaxed by slow 
annealing of both ionic and electronic degrees of freedom. 

The formation energies were evaluated using the 
\texttt{SIESTA} code \cite{Soler02} by relaxing the 
ionic coordinates and the in-plane cell dimensions. 
The same norm-conserving pseudopotentials and density 
functional as in the plane wave calculations together with an optimized 
double-$\zeta$ plus polarization function (DZP) basis set were used.
A 2$\times$2$\times$2 k-point grid (including the $\Gamma$ point) 
was employed in order to 
obtain accurate defect formation energies.\cite{VandeWalle04} 

\section{\label{sec:results}RESULTS AND DISCUSSION}

\subsection{Off-plane recoils}

The outcomes of our simulations are summarized in Figure~\ref{fig:char1}
(movies of selected MD trajectories are available online
\cite{EPAPS-note}).
We first discuss the simulation results for the off-plane displacements
($\phi$$\in$$\left\{0^\circ; 30^\circ; 60^\circ \right\}$) of carbon atoms in inequivalent positions $\alpha$ and $\beta$.
The outcomes can be divided into four major classes: (i) no defect formation due
to insufficient transferred momentum or due to instantaneous recombination of 
the recoil atom with the vacancy ('n'); 
(ii) separated interstitial-vacancy pairs 
(I-V); (iii) intimate interstitial-vacancy
pairs (iIV); (iv) displacement cascades ('c') in which the recoil atom 
is able to displace other atoms in the lattice.
The latter case can be viewed 
as a series of elementary events of types (i)--(iii).
The simulation of displacement cascades is beyond the scope of this study
and would require a larger unit cell than the one used here.

\begin{figure}
\includegraphics[width=8.6cm]{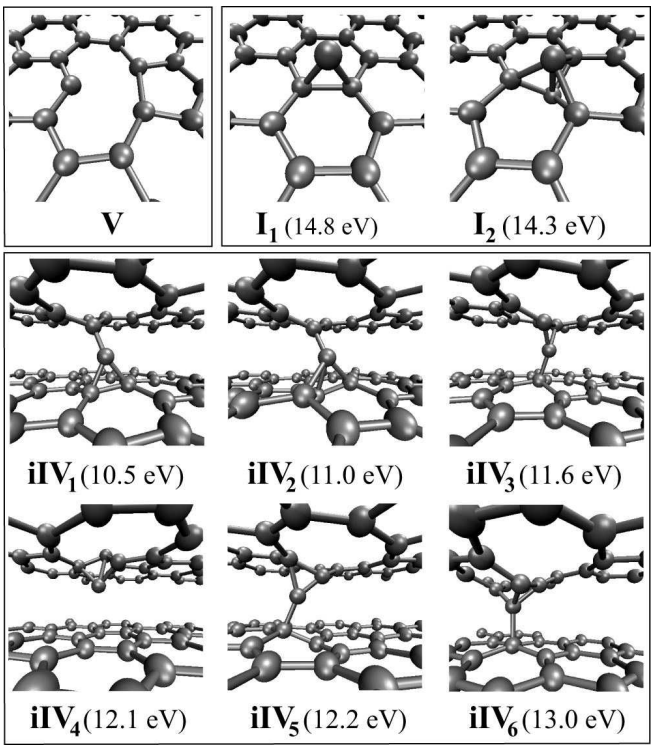}
\caption{\label{fig:eps1}  
Perspective views of the atomic structures 
of the vacancy (top, left), interstitial (top, right), and 
intimate Frenkel pair (bottom) defects observed in our simulations. 
The formation energies are given in parentheses. The values given for interstitial 
defects refer to the formation energies of corresponding Frenkel pairs.
For iIV defects the created vacancy is situated in the upper graphene layer.}
\end{figure}

\begin{figure}
\includegraphics[width=8.6cm]{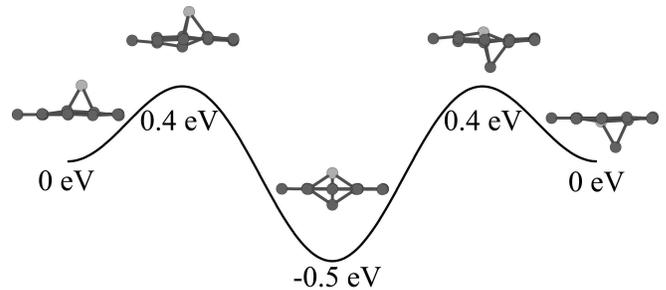}
\caption{\label{fig:eps4}  
The scheme of the self-diffusion process in graphite along the $c$-axis. The diffusing 
carbon adatom (highlighted in the figure) substitutes one of the carbon atoms in the graphite layer.
The relative energies of local minima and transition states are shown.}
\end{figure}

The formation of well-separated Frenkel pairs was observed
for atoms in both $\alpha$ and $\beta$ positions at $T$$\ge$25~eV.
Surprisingly, the interstitial defects were produced only
in the form of a symmetric ``dumbell'' structure (I$_2$) \cite{Li05,Ma05}
where the two carbon atoms are symmetrically displaced 
from the graphene plane (Fig.~\ref{fig:eps1}, top). 
Despite the highly distorted coordination sphere of these atoms,
the C--C distance of 1.58~\AA\ is close to the one of a typical $\sigma$ bond. 
The core atomic structure is the same as for the
[1.1.1]propellane molecule for which a very similar C--C bond length 
(1.60$\pm$0.02~\AA) has been observed experimentally.\cite{Wiberg85}
No single off-plane recoil led to the ``bridge'' 
structure (I$_1$) \cite{Nordlund96,Lehtinen03,Li05} with the interstitial
atom situated between two graphene planes.
The formation energy of I$_2$ ($E_f$=14.3~eV, the value refers to the formation energy of the corresponding I-V pair) is only 0.5~eV lower than the one of I$_1$ ($E_f$=14.8~eV) where the ``bridge'' interstitial defect is bonded only to
the neighbor atom in the same layer. 
In this case, a steric repulsion with the opposite 
graphene layer contributes to the destabilization of the I$_1$ defect.
However, bonding to the opposite layer leads to more stable 
shared interstitial defect structures.\cite{Telling03,Li05}
In the ylid ($E_f$=14.1~eV) and spiro ($E_f$=13.1~eV) configurations,
the shared interstitial atom is additionally 
bound to one and, respectively, to two carbon atoms of the
adjacent layer.
These structures have not been observed in our MD simulations.
The observed preference for the I$_2$ configuration in graphite 
may have the following origin.
In the ``dumbell'' configuration the recoil atom is able to 
transfer its excess kinetic energy to the other atoms more efficiently
than in the case of the ``bridge'' configuration. 
At the same time, the formation of shared interstitials 
requires the improbable collective motion of a number of atoms in the two 
adjacent graphene layers in the direction of the recoil atom.
This explains the observed high probability for the formation of the 
I$_2$ defect structure in the early stages of the 
radiation-induced defect formation. 
For the isolated graphene sheet, I$_1$ is 0.2~eV
more stable than I$_2$ due to the absence of the steric repulsion with 
the adjacent graphene layer. 
In curved graphenic structures, like carbon nanotubes,
the ``bridge'' interstitial defect undergoes further stabilization.  
Our first principles calculations predict that
the transition from the defect structure I$_2$ to 
the structure I$_1$ in graphite
is characterized by an activation barrier of 0.9~eV. 
The ``dumbell'' interstitial 
can also be viewed as a stable intermediate of the self-diffusion process 
in graphite along the $c$-axis, occurring via 
the substitution of a carbon atom in the graphene layer.\cite{Xu93} 
The energy diagram for the diffusion process along the $c$-axis is shown in Figure~\ref{fig:eps4}.

Formation of intimate interstitial-vacancy pairs (iIV) requires lower 
transferred kinetic energies. 
At $T$=20~eV we observed the formation of two low energy iIV pairs, iIV$_1$ ($E_f$=10.5~eV) and iIV$_2$
($E_f$=11.0~eV) (Fig.~\ref{fig:eps1}, bottom).
The displaced atom bridges the defect vacancy with two, respectively, three 
neighbor atoms in the opposite layer, which undergo rehybridization.
A fine scan of the transferred
momentum space indicates a $T_d$ value of 18~eV 
for graphite, in agreement with other reported values.\cite{Banhart97,Smith01}
As a consequence, the use of a particle beam energy capable of achieving a maximum 
kinetic energy transfer just above $T_d$ will {\it selectively} create iIV defects.
This value for $T_d$ would correspond to the maximum kinetic energy transferred by an
electron beam of 90~keV.\cite{Smith01}
In the case of carbon nanotubes, $T_d$ is expected to be lower due 
to curvature effects.\cite{Urita05}
This proves the crucial role of the iIV defects 
in the reinforcement of carbon nanotube bundles \cite{Kis04,daSilva05} 
produced by 80~keV electron irradiation.
Our results suggest the optimal conditions for the modification of
mechanic and electronic properties of carbon-based layered
nanostructures by means of the formation of iIV defects. 
For graphite and closely related nanostructures, electron beam acceleration voltages of 90--110~kV
can be used.
Such modifications are non-destructive since iIV defects tend to self-recombine
without producing extensive damage of the nanostructure.\cite{Urita05}
This is also supported by the fact that the barriers for iIV$_1$ 
defect recombination \cite{Ewels03} and for the transformation of iIV$_1$ into iIV$_2$ 
(0.9~eV in this study) lie below the formation energies of I-V pairs.

Our computed formation energy for the previously proposed iIV$_1$ structure \cite{Ewels03}
is in good agreement with the values reported in 
other studies.\cite{Ewels03,Li05,daSilva05} 
However, MD simulations on a longer time scale 
indicate that the asymmetric iIV$_1$ defect in graphite 
is not stable against recombination at 350~K if the shear of 
neighboring graphite layers is allowed. 
By contrast, the symmetric iIV$_2$ defect is
stable throughout our MD simulations.
Two other intimate Frenkel pairs, iIV$_3$ ($E_f$=11.6~eV) and iIV$_5$ ($E_f$=12.2~eV)
have been obtained upon off-plane recoils caused by 
larger transferred momenta. 
In both structures the displaced carbon atom
is linked to two carbon atoms in its host layer and one atom in 
the neighboring layer.      
It is notable that the formation of
iIV defects has been observed {\it only} upon recoil of the $\beta$ carbon atom.
We explain this observation by the large probability of 
instantaneous recombination of the $\alpha$ atom recoils due to the local
arrangement of atoms in the adjacent layer.  

\subsection{In-plane recoils}

\begin{figure}
\includegraphics[width=8.6cm]{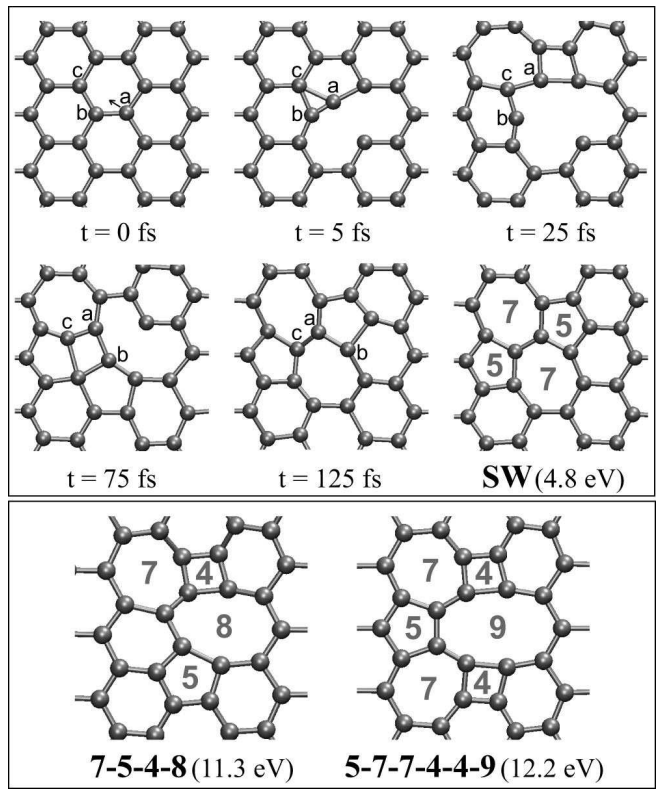}
\caption{\label{fig:eps2}  
Top: The mechanism of formation of a Stone-Wales defect 
upon in-plane knock-on displacement ($T$=30~eV). 
The carbon atoms involved in the rearrangement are marked with letters.
Bottom: Atomic structures of 
7-5-4-8 and 5-7-7-4-4-9 topological defects (arabic numbers indicate non-sixmembered rings).}
\end{figure}

The formation of defects after displacement in the graphene
plane ($\phi$=90$^\circ$) requires higher transferred energies $T$$\ge$30~eV.
At $T$=30~eV ($\theta$=30$^\circ$) we observed the formation of a Stone-Wales (SW) 
defect,\cite{Stone86} 
which is the lowest energy ($E_f$=4.8~eV) defect in graphite. 
The mechanism of its formation involves the cyclic permutation 
of three carbon atoms occurring during the first 100~fs after the knock-on 
collision (Fig.~\ref{fig:eps2}, top).  
A much lower activation barrier of $\approx$10~eV is required 
when the SW defect is formed upon simultaneous in-plane rotation of two neighboring carbon atoms.\cite{Kaxiras88} 
However, this mechanism {\it cannot} be realized upon knock-on collisions because in this case 
the kinetic energy is transferred to a single atom. 
Irradiation of graphene-based materials, using an electron beam of energy just above
150~keV and oriented along the graphene plane, will result in an {\it increase}
of yield of SW defects. 
This can be used for tuning electronic properties of materials.\cite{Crespi97} 
However, because of the high energy transfer required for their 
formation, SW defects will be accompanied by the formation 
of Frenkel pairs, which form upon low energy ($T$$<$30~eV) off-plane recoils.

For $T$$>$30~eV two possible general mechanisms of defect 
formation have been identified. The first
one involves the formation of strained structures containing 
non-sixmembered rings, which have formation energies higher 
than the formation energy of the SW defect. 
Two such structures, 7-5-4-8 ($E_f$=11.3~eV) and 5-7-7-4-4-9 ($E_f$=12.2~eV)
have been observed in our simulations (Fig.~\ref{fig:eps2}, bottom).
The second mechanism involves the expulsion of one carbon atom 
from the graphene plane shortly after the collision. 
In this case interstitial-vacancy pairs are formed.
We observed formation of the ``bridge'' interstitial defect 
I$_1$ caused by the expulsion of a carbon atom with low kinetic energy. 
In addition, two new intimate
interstitial-vacancy pairs, iIV$_4$ ($E_f$=12.1~eV) and iIV$_6$ ($E_f$=13.0~eV)
have been characterized. 
In the iIV$_4$ structure the defect is localized in the graphene
layer where the collision took place. 
On the contrary,
the displaced carbon atom in the iIV$_6$ structure bridges three atoms of its
host layer with one of the neighboring layers. The formation
energies of all six iIV structures found in our 
simulations lie within a narrow interval of 2.5~eV, 
and they are all below the formation energies of separated I-V pairs.   
These defects should be stable at long time scales and at moderate temperatures.

\section{\label{sec:conclusions}CONCLUSIONS}

In conclusion, we performed an \textit{ab initio} molecular dynamics study 
of radiation-induced defect formation in graphite.
A variety of different defects, including structures which have 
never been discussed previously, were observed in our simulations. 
The produced defects depend strongly upon the direction and magnitude of the transferred momentum, resulting in the selective formation 
of certain defect structures. 
We showed the crucial role played by the early stage dynamics in the defect formation process, 
and we identified the conditions at which selective creation of defects can be achieved.   
In particular, we identified an interval of electron beam energies
at which only low-energy intimate Frenkel pair defects bridging adjacent graphene layers
are produced.  We also conclude that Stone-Wales defects, characterized by the lowest formation energy, 
cannot be produced selectively upon irradiation.   
Our results are of practical importance for radiation-assisted 
manufacturing of carbon materials and nanostructures with new desired properties
and functions. 

\section*{ACKNOWLEDGMENTS}

The authors acknowledge L.~Forr\'o, A.~Kis, A.~Kulik, S.~Reich, 
B.~I.~Yakobson, and A.~Zettl for discussions.
O.~Y. thanks the Swiss NSF for financial support. 
The computational resources were provided by the CSCS and the DIT-EPFL.

\end{document}